\documentclass[english]{article}
\usepackage[T1]{fontenc}
\usepackage[latin9]{inputenc}
\usepackage{textcomp}
\usepackage{amsmath}
\usepackage{amssymb}
\usepackage{esint}

\makeatletter

\newcommand{\noun}[1]{\textsc{#1}}

\newcommand{\lyxaddress}[1]{
\par {\raggedright #1
\vspace{1.4em}
\noindent\par}
}

\makeatother

\usepackage{babel}
\begin{document}

\title{Metriplectic Algebra\\
for Dissipative Fluids\\
in Lagrangian Formulation%
}

\author{Massimo Materassi}

\maketitle

\lyxaddress{Istituto dei Sistemi Complessi, ISC-CNR, via Madonna del Piano 10,
50019 Sesto Fiorentino (Florence), Italy, e-mail: massimo.materassi@isc.cnr.it}
\begin{abstract}
It is known that the dynamics of dissipative fluids in Eulerian variables
can be derived from an algebra of Leibniz brackets of observables,
the metriplectic algebra, that extends the Poisson algebra of the
zero viscosity limit via a symmetric, semidefinite component. This
metric bracket generates dissipative forces.

The metriplectic algebra includes the conserved total Hamiltonian
$H$, generating the non-dissipative part of dynamics, and the entropy
$S$ of those microscopic degrees of freedom draining energy irreversibly,
that generates dissipation. This $S$ is a Casimir of the Poisson
algebra to which the metriplectic algebra reduces in the frictionless
limit.

In the present paper, the metriplectic framework for viscous fluids
is re-written in the Lagrangian Formulation, where the system is described
through material variables: this is a way to describe the continuum
much closer to the discrete system dynamics than the Eulerian fields.
Accordingly, the full metriplectic algebra is constructed in material
variables, and this will render it possible to apply it in all those
cases in which the Lagrangian Formulation is preferred.

The role of the entropy $S$ of a metriplectic system is as paramount
as that of the Hamiltonian $H$, but this fact may be underestimated
in the Eulerian formulation because $S$ is not the only Casimir of
the symplectic non-canonical part of the algebra. Instead, when the
dynamics of the non-ideal fluid is written through the parcel variables
of the Lagrangian formulation, the fact that entropy is symplectically
invariant appears to be more clearly related to its dependence on
the microscopic degrees of freedom of the fluid, that do not participate
at all to the symplectic canonical part of the algebra (which, indeed,
involves and evolves only the macroscopic degrees of freedom of the
fluid parcel).
\end{abstract}

\section{Introduction\label{sec:Introduction}}

The history of Theoretical Physics is, to a certain extent, that of
the discovery of symmetries of physical laws, allowing to bypass the
necessity of solving the equations of motion explicitly and gaining
deep insights about the essence of first principles themselves.

The highest achievements of this simplification process are the \emph{least
action principles} \cite{LAP.Maupertuis,LAP.Hamilton}, with the \emph{Feynman
path integral} \cite{Feynman.Hibbs} as their most recent descendant,
and the study of \emph{invariances} \cite{Noether.1-2}; the \emph{Hamiltonian
formalism} \cite{LAP.Hamilton,Goldstein.book} and the \emph{Hamilton-Jacobi
theory} \cite{Goldstein.book,Hamilton.HJ}. In the context of Hamiltonian
mechanics, the dynamics of physical systems appears in the form of
an \emph{algebra of Poisson brackets} \cite{Arnold.book}, composing
together the physical observables to both represent the motion of
the system and the symmetry properties of its dynamics. This \emph{route
to the algebrization} of dynamics also leads to Dirac's formulation
of Quantum Mechancis \cite{Dirac.book}, according to which the algebra
of quantum observables is simply a commutation algebra of operators,
isomorphic to the Poisson algebra of the respective classical ones.

However, almost all the benefits of the just mentioned development
are, generally speaking, restricted to the Physics of \emph{non dissipative
systems}: in Lagrangian and Hamiltonian mechanics, as well as in the
context of action principles \cite{Goldstein.book,Arnold.book}, only
systems undergoing conservative forces are treated, while no form
of ``dissipation'' is considered in the fundamental quantum laws,
intended as the basic principles of Physics \cite{Weinberg.book.1}.
When dissipative quantum systems are referred to, one typically considers
open systems in interaction with some ``environment'' only partially
observed \cite{Tarasov.book}, and these are not regarded as ``fundamental''
(one should however mention the ``line of thought'', expressed in
\cite{PrigoPetrovsky} for instance, in which dissipation is included
in the fundamental laws of Quantum Mechanics).

A fruitful attempt to meet dissipative systems along the route to
algebrization of Physics is represented by the use of \emph{Leibniz
algebræ} \cite{Gorba.Leibniz,Guha.Leibniz} generalizing the symplectic
product. The dynamics of \emph{energetically closed systems relaxing
to asymptotic equilibria due to dissipation} has been described in
this way by Morrison in \cite{Morrison.bracket.formulation.84} via
the so called \emph{metriplectic formalism}. The non-dissipative limit
of the system is Hamiltonian, so that there exists some function $H$
and an algebra of Poisson brackets that describes the system in the
absence of dissipation. When dissipation is turned on, the Hamiltonian
is still constant during the motion, but friction drives the system
to an asymptotic equilibrium: this is done generalizing the Poisson
bracket so to include a symmetric semidefinite component, referred
to as \emph{metric bracket}. Then, dissipative processes are generated
through that symmetric extension by an observable representing the
entropy $S$ of the closure of the system.

Quite a few dynamical systems have been reformulated as metriplectic:
in \cite{Morrison.bracket.formulation.84} the kinetic Vlasov-Poisson
approximation of a collisional plasma was described as Hamiltonian
in the collisionless limit, while collisional terms are shown to arise
from a metric bracket. In \cite{Morrison.84.001} a non-ideal fluid
described in Eulerian variables (EV) is presented as a non-canonical
Hamiltonian system, with the addition of a metric bracket providing
the dissipative terms due to the finite viscosity and thermal conductivity.
In \cite{Morrison.paradigm.86} the non-canonical Hamiltonian dynamics
of a free rigid body throughout the space of its angular velocity
is enriched by a metric contribution through which the rotator is
made relax down to asymptotic equilibria at which the system spins
around one of its inertial axes. In the same work, the dissipative
Vlasov-Poisson equation is examined again.

A general review of Poisson and metric brackets to describe energetically
isolated or non-isolated systems (referred to as \emph{complete} and
\emph{incomplete}) may be found in \cite{Morrison.2009}.

In his PhD thesis \cite{Fish.th.PhD}, Fish examines metriplectic
systems of various types under the point of view of manifold properties,
and also gives interesting examples from applied physics and biophysics.

The metriplectic system describing neutral fluids in \cite{Morrison.84.001}
has been generalized to non-ideal magneto-hydrodynamics in \cite{Mate.Tassi.metriplectic.MHD},
while examples of how to algebrize simple mechanical systems with
friction are provided in \cite{algebrizing.friction}.

In the present paper,\emph{ the Lagrangian Formulation (LF) of the
metriplectic algebra for a viscous fluid} is constructed. The symplectic
part of the metriplectic system is taken from \cite{Morrison.84.001,PadMorr.01,PadMorr.02},
while the metric part is an original contribution presented here for
the first time, as far as the author is aware of, \emph{by mapping
the metric bracket in EV to its expression in parcel variables}.

Even if rather interesting from the viewpoint of mathematical completeness,
still the translation of the Eulerian metriplectic algebra to the
Lagrangian one can be questioned to be worth the effort in physical
terms. Instead, it should be underlined that the symmetry-related
role of the fluid entropy appears much clearer in the Lagrangian algebra
than in the Eulerian one, not to mention that whenever the use of
LF is preferred to that of EF, the expressions found here will be
applied.

The fluid entropy has zero Poisson bracket with any other quantity
in both formulations, but the expression of the symplectic product
in Lagrangian variables makes it clear that $S$ is not a Casimir
invariant due to the parcel relabeling symmetry (that allows the fluid
to possess an Eulerian representation at all), but simply because
it encodes of degrees of freedom involved in parcel dynamics only
through dissipation.

In Section \ref{sec:Metriplectic-complete-systems} the general framework
of metriplectic complete systems is sketched, while in Section \ref{sec:Casimir-Invariants}
we discuss briefly the role of Casimir invariants of the theory with
respect to algebra reduction and dissipative processes.

Section \ref{sec:LF-viscous} is dedicated to the key result of this
paper: the Lagrangian Formulation is constructed for viscous fluids,
and their metriplectic algebra is formulated in the material variables.
With this result in mind, a speculation on the nature of the fluid
entropy as a Casimir invariant of the theory is presented in Section
\ref{sec:Entropy-Casimir}.

Conclusions are reported in Section \ref{sec:Conclusions}, where
possible applications and future developments of the present research
are also sketched.

\section{Metriplectic complete systems\label{sec:Metriplectic-complete-systems}}

Consider an energetically closed system with dissipation, and describe
its state as a point $\psi$ moving in a suitable phase space $\mathbb{V}$.
Also, refer to its algebra of observables $\mathcal{O}$ as a subset
of $C^{\infty}\left(\mathbb{V},\mathbb{R}\right)$.

According to the metriplectic scheme, its dynamics $\dot{\psi}$ will
be expressed as the sum of a non-dissipative part $\dot{\psi}_{\mathrm{non-diss}}=\left\{ \psi,H\right\} $,
generated by the Hamiltonian $H\in\mathcal{O}$ through a Poisson
bracket structure, and the dissipative part $\dot{\psi}_{\mathrm{diss}}=\lambda\left(\psi,S\right)$,
where $\left(.,.\right)$ is a \emph{symmetric semidefinite Leibniz
bracket}
\[
\begin{array}{cccc}
\left(A,B\right)=\left(B,A\right), & \left(A,A\right)\leq0 & \forall & A,B\in\mathcal{O},\end{array}
\]
referred to as \emph{metric bracket}, and $\lambda$ is a negative
constant parameter (making physical sense only in the correspondence
of the asymptotic equilibrim \cite{Mate.Tassi.metriplectic.MHD}).
The generator $S$ of the dissipative dynamics $\dot{\psi}_{\mathrm{diss}}$
has zero Poisson bracket with any other observable depending on $\psi$
\begin{equation}
\begin{array}{ccc}
\left\{ S,A\right\} =0 & \forall & A\in\mathcal{O},\end{array}\label{eq:S.Casimir}
\end{equation}
while the metric bracket $\left(.,.\right)$ must have $H$ among
its null modes
\begin{equation}
\begin{array}{ccc}
\left(H,A\right)=0 & \forall & A\in\mathcal{O}.\end{array}\label{eq:H.metric.invariant}
\end{equation}

If the evolution of the system works as
\begin{equation}
\dot{\psi}=\dot{\psi}_{\mathrm{non-diss}}+\dot{\psi}_{\mathrm{diss}}=\left\{ \psi,H\right\} +\lambda\left(\psi,S\right),\label{eq:psi.dot}
\end{equation}
then any quantity $\Phi\in\mathcal{O}$ depending on $\psi$ evolves
according to the same rule
\begin{equation}
\dot{\Phi}\left(\psi\right)=\left\{ \Phi\left(\psi\right),H\right\} +\lambda\left(\Phi\left(\psi\right),S\right).\label{eq:Phi.dot}
\end{equation}
Due to the conditions (\ref{eq:S.Casimir}) and (\ref{eq:H.metric.invariant}),
this general rule also implies
\begin{equation}
\begin{array}{cc}
\dot{H}=0, & \dot{S}=\lambda\left(S,S\right)\ge0:\end{array}\label{eq:H.dot.S.dot}
\end{equation}
the first of these equations means that $H$ is constant because it
is not altered by dissipation, that just redistributes energy but
does not destroy it; the second condition in (\ref{eq:H.dot.S.dot})
states that $S$ asymptotically and monotonically grows during the
motion, as a Lyapunov quantity does in the correspondence of an asymptotic
stable state \cite{prigogine.chaos.stochasticity}.

The conditions (\ref{eq:S.Casimir}) and (\ref{eq:H.metric.invariant}),
together with the properties of $\left\{ .,.\right\} $ and $\left(.,.\right)$
as Leibniz brackets \cite{Gorba.Leibniz}, allow for the definition
of a total metriplectic generator $F=H+\lambda S$ so that, provided
the new bracket
\begin{equation}
\left\langle \left\langle A,B\right\rangle \right\rangle =\left\{ A,B\right\} +\left(A,B\right)\label{eq:metriplectic.bracket.def}
\end{equation}
is defined, one may simply state
\[
\dot{\psi}=\left\langle \left\langle \psi,F\right\rangle \right\rangle ,
\]
and $\dot{\Phi}=\left\langle \left\langle \Phi,F\right\rangle \right\rangle $
for any observable $\Phi$. The new Leibniz structure defined in (\ref{eq:metriplectic.bracket.def})
is the \emph{metriplectic bracket}, while the metriplectic generator
$F$ is sometimes referred to as \emph{free energy}.

\section{Casimir Invariants\label{sec:Casimir-Invariants}}

The condition (\ref{eq:S.Casimir}) attributes to the metric generator
$S$, whatever it is physically, the algebraic character of \emph{Casimir
invariant} (CI) of the Poisson bracket $\left\{ .,.\right\} $.

Now, the metric generator may be a CI for one of the two following
reasons.

Either, the symplectic bracket $\left\{ .,.\right\} $ includes derivatives
with respect to the variables on which $S$ depends too, and nevertheless
admits a non-trivial kernel to which $S$ belongs; this case, we will
refer to as C1, is typical for Poisson algebrae $\left(\mathcal{A}_{\mathrm{red}},\left\{ .,.\right\} _{\mathrm{red}}\right)$
obtained by reducing some Poisson algebra $\left(\mathcal{A},\left\{ .,.\right\} \right)$
to the algebra of all the observables invariant under a certain group
of transformations $\mathbb{G}$: if $G\in\mathcal{A}$ is a symplectic
generator of those transformations, clearly $\left\{ \Phi,G\right\} _{\mathrm{red}}=0$
for any element $\Phi\in\mathcal{A}_{\mathrm{red}}$, so that $G$
is a CI for the bracket $\left\{ .,.\right\} _{\mathrm{red}}$.

Or, the variables forming $S$ do not appear at all in the definition
of the Poisson bracket, so that $S$ belongs to the kernel of it as
does any variable outside the system; this other case, referred to
as C2, is that of a Poisson algebra $\left(\mathcal{A}_{0},\left\{ .,.\right\} _{0}\right)$
describing a system of variables $\psi_{0}$ in interaction with some
environment, of which an effective description is given via a variable
$z$ external to the system: then, any $C\left(z\right)$ is \emph{trivially}
a CI of $\left\{ .,.\right\} _{0}$, since the latter depends only
on derivatives with respect to $\psi_{0}$ but does not involve any
derivative in $z$. A metriplectic system describing the relaxation
of ``macroscopic'' variables $\psi_{0}$ due to the interaction
with some \emph{microscopic degrees of freedom} ($\mu$DoF) may be
conceived by defining a metric bracket ``driven'' by $C\left(z\right)$
and acting on the ``total'' state $\psi\overset{\mathrm{def}}{=}\left(\psi_{0},z\right)$.

Throughout the literature mentioned in Section \ref{sec:Introduction},
one meets examples of both kinds C1 and C2.

The free dumped rotator presented in \cite{Morrison.paradigm.86},
and revisited in \cite{Fish.th.PhD} is easily recognized to be a
C1 case: the square angular momentum is such a CI when the phase space
of the rotator is reduced from the 6 dimensional space of angles and
their canonical momenta to the $\mathbb{R}^{3}$ of angular velocities.
Systems with dissipative constants regarded as control parameters
depending on an external variable are properly C2 cases (e.g., the
Lodka-Volterra, Lorentz and Van Der-Pole systems in \cite{Fish.th.PhD},
or the elementary mechanics dissipative systems reported in \cite{algebrizing.friction},
where the external variable is the state of a thermal bath). Last
but not least, deciding whether the Boltzmann entropy playing the
role of the metric generator for the Vlasov-Poisson collisional plasma
is a C1 or C2 quantity deserves a deeper investigation, involving
the fact that Vlasov-Poisson equation results from the truncation
of a \emph{hierarchy of equations} involving many-particle variables
\cite{Balescu.book}, the symplectic limit of which has been studied
in \cite{Morrison.BBGKY}. The origin of being a CI for the entropy
of a viscous fluid is investigated here, writing the Lagrangian metriplectic
algebra explicitly, as done in Section \ref{sec:LF-viscous} below.

\section{Lagrangian Formulation for viscous fluids\label{sec:LF-viscous}}

In \cite{Morrison.84.001} the viscous fluid equation is described
in the Eulerian Formalism (EF), via the fields mass density $\rho\left(\vec{x},t\right)$,
velocity $\vec{v}\left(\vec{x},t\right)$ and mass-specific entropy
density $\sigma\left(\vec{x},t\right)$. In the non-dissipative limit
the dynamics takes a non-canonical Hamiltonian form: the Poisson bracket
between two any functionals $\Phi\left[\rho,\vec{v},\sigma\right]$
and $\Psi\left[\rho,\vec{v},\sigma\right]$ is defined as
\begin{equation}
\begin{array}{l}
\left\{ \Phi,\Psi\right\} _{E}=-\int\limits _{\mathbb{R}^{3}}d^{3}x\left[\dfrac{\delta\Phi}{\delta\rho}\partial_{\alpha}\left(\dfrac{\delta\Psi}{\delta v_{\alpha}}\right)+\dfrac{\delta\Psi}{\delta\rho}\partial_{\alpha}\left(\dfrac{\delta\Phi}{\delta v_{\alpha}}\right)-\dfrac{1}{\rho}\dfrac{\delta\Phi}{\delta v_{\alpha}}\epsilon_{\alpha\gamma\beta}\epsilon^{\beta\delta\eta}\dfrac{\delta\Psi}{\delta v_{\gamma}}\partial_{\delta}v_{\eta}+\right.\\
\\
\left.+\dfrac{1}{\rho}\partial_{\alpha}\sigma\left(\dfrac{\delta\Phi}{\delta\sigma}\dfrac{\delta\Psi}{\delta v_{\alpha}}-\dfrac{\delta\Psi}{\delta\sigma}\dfrac{\delta\Phi}{\delta v_{\alpha}}\right)\right],
\end{array}\label{eq:Poisson.Bracket.in.EV}
\end{equation}
where Greek indices are used for the $SO\left(3\right)$-vector components
of $\vec{v}$ and of the position $\vec{x}$ in the space and a summation
convention holds, so that scalar products in $\mathbb{R}^{3}$ read
$\vec{v}\cdot\vec{w}=v^{\alpha}w_{\alpha}$. The symbol $\partial_{\alpha}=\frac{\partial}{\partial x^{\alpha}}$
is used for spatial gradients.

The Hamiltonian functional of the system reads
\begin{equation}
H\left[\rho,\vec{v},\sigma\right]=\int\limits _{\mathbb{R}^{3}}d^{3}x\left[\dfrac{\rho v^{2}}{2}+\rho U\left(\rho,\sigma\right)+\rho\phi\right],\label{eq:H.EV}
\end{equation}
where $\rho Ud^{3}x$ is the amount of internal energy attributed
to the infinitesimal volume $d^{3}x$ around the position $\vec{x}$.
$\varphi$ is an external potential.

$H$ generates the motion of any observable $\Phi\left[\rho,\vec{v},\sigma\right]$
as $\dot{\Phi}=\left\{ \Phi,H\right\} _{E}$ thanks to the Poisson
bracket (\ref{eq:Poisson.Bracket.in.EV}): the \emph{non-dissipative
Navier-Stokes equations}
\[
\begin{cases}
\partial_{t}v_{\alpha}=-v_{\beta}\partial^{\beta}v_{\alpha}-\dfrac{1}{\rho}\partial_{\alpha}p-\partial_{\alpha}\varphi,\\
\\
\partial_{t}\rho=-\partial^{\alpha}\left(\rho v_{\alpha}\right),\\
\\
\partial_{t}\sigma=-v_{\alpha}\partial^{\alpha}\sigma,
\end{cases}
\]
where $p$ is the pressure, hence follow.

Let then viscosity and thermal conductivity be finite.

Let the viscosity tensor be of the form $\Sigma^{\alpha\beta}=\Lambda^{\alpha\beta\gamma\delta}\partial_{\gamma}v_{\delta}$,
with $\Lambda$ constant; let the heat flux $\vec{I}$ be related
to the local temperature $T$ as $I_{\beta}=-\kappa\partial_{\beta}T$:
then, the symplectic algebra (\ref{eq:Poisson.Bracket.in.EV}) must
be completed by the metric bracket
\begin{equation}
\begin{array}{l}
\left(\Phi,\Psi\right)_{E}=\dfrac{1}{\lambda}\int\limits _{\mathbb{R}^{3}}d^{3}x\left\{ T\Lambda_{\alpha\beta\gamma\delta}\left[\partial^{\alpha}\left(\dfrac{1}{\rho}\dfrac{\delta\Phi}{\delta v_{\beta}}\right)-\dfrac{1}{\rho T}\partial^{\alpha}v^{\beta}\dfrac{\delta\Phi}{\delta\sigma}\right]\left[\partial^{\gamma}\left(\dfrac{1}{\rho}\dfrac{\delta\Psi}{\delta v_{\delta}}\right)-\dfrac{1}{\rho T}\partial^{\gamma}v^{\delta}\dfrac{\delta\Psi}{\delta\sigma}\right]+\right.\\
\\
\left.+\kappa T^{2}\partial^{\alpha}\left(\dfrac{1}{\rho T}\dfrac{\delta\Phi}{\delta\sigma}\right)\partial_{\alpha}\left(\dfrac{1}{\rho T}\dfrac{\delta\Psi}{\delta\sigma}\right)\right\} 
\end{array}\label{eq:metric.bracket.EV}
\end{equation}
so that, given the total entropy of the fluid as
\[
S\left[\rho,\sigma\right]=\int_{\mathbb{R}^{3}}\rho\sigma d^{3}x,
\]
the dynamics reads
\[
\dot{\Phi}=\left\{ \Phi,H\right\} _{E}+\lambda\left(\Phi,S\right)_{E},
\]
giving rise to the non-ideal equations of motion:
\begin{equation}
\begin{cases}
\partial_{t}v_{\alpha}=-v_{\beta}\partial^{\beta}v_{\alpha}-\dfrac{1}{\rho}\partial_{\alpha}p-\partial_{\alpha}\varphi+\dfrac{1}{\rho}\partial^{\kappa}\left(\Lambda_{\kappa\alpha\beta\gamma}\partial^{\beta}v^{\gamma}\right),\\
\\
\partial_{t}\rho=-\partial^{\alpha}\left(\rho v_{\alpha}\right),\\
\\
\partial_{t}\sigma=-v_{\alpha}\partial^{\alpha}\sigma+\dfrac{1}{\rho T}\Lambda_{\kappa\alpha\beta\gamma}\partial^{\kappa}v^{\alpha}\partial^{\beta}v^{\gamma}+\dfrac{\kappa}{\rho T}\partial^{2}T.
\end{cases}\label{eq:NS.viscosa.EV}
\end{equation}
The symbol $\partial^{2}=\partial^{\alpha}\partial_{\alpha}$ has
been used.

In the LF, the fluid is subdivided into material parcels labeled by
a continuous three-index $\vec{a}$, and the motion and evolution
of each $\vec{a}$-th parcel is followed \cite{Bennett.book}. As
far as its motion throughout the space is concerned, the $\vec{a}$-th
fluid parcel is described at time $t$ by its position $\vec{\zeta}\left(\vec{a},t\right)$
and its momentum $\vec{\pi}\left(\vec{a},t\right)$ (in order to give
a more concrete sense to the label $\vec{a}$, the choice
\begin{equation}
\vec{a}=\vec{\zeta}\left(\vec{a},0\right)\label{eq:a.initial.zeta}
\end{equation}
can be made). Since the parcel is a system of $\mathbb{O}\left(10^{23}\right)$
microscopic particles, it must be equipped also by some variable describing
those $\mu$DoF: its mass-specific entropy density $s\left(\vec{a},t\right)$
is given this role \cite{subfluid.mate.tassi.conso}. The fact that
the $\mu$DoF of the $\vec{a}$-th parcel are all encoded in the thermodynamical
variable $s\left(\vec{a},t\right)$ suggests that they are treated
\emph{statistically}. In a sense, the metriplectic formalism is the
algebrization of a stochastic dynamics in which what remains of the
probabilistic noise is its equilibrim thermodynamics \cite{algebrizing.friction}.

A vivid representation of the variables $\left(\vec{\zeta},\vec{\pi},s\right)$
may be that $\left(\vec{\zeta},\vec{\pi}\right)$ are the variables
of the parcel's centre-of-mass, while $s$ encodes the thermodynamics
of the relative variables \cite{MateLusa}.

The field configuration $\left(\vec{\zeta},\vec{\pi},s\right)$ represents
the state of the fluid in LF, let's indicate its functional phase
space as $\mathbb{V}_{L}$. In LF the hypothesis of parcel identity
conservation is made: this means that at every time $t$ the map $\vec{a}\mapsto\vec{\zeta}\left(\vec{a},t\right)$
is a diffeomorphism from the space initially occupied by the continuum
$\mathbb{D}_{0}$ and the one it occupies at time $t$, $\mathbb{D}\left(t\right)\subseteq\mathbb{R}^{3}$.
If its Jacobian matrix $J_{i}^{\mu}=\frac{\partial\zeta^{\mu}}{\partial a^{i}}$
is defined, with the volume expansion factor $\mathcal{J}=\det J$,
then the measure of the infinitesimal volume $d^{3}\zeta\left(\vec{a},t\right)$
of the $\vec{a}$-th parcel at time $t$ is related to its initial
volume $d^{3}a$ by the law $d^{3}\zeta=\mathcal{J}d^{3}a$. Also,
these diffeomorphisms show a (semi)-group property with respect to
the parameter $t$: $\vec{\zeta}\left(\vec{a},t_{1}+t_{2}\right)=\vec{\zeta}\left(\vec{\zeta}\left(\vec{a},t_{1}\right),t_{2}\right)$.

Vector components of $\vec{\zeta}$ and $\vec{\pi}$ are labeled by
Greek indices, as $\vec{v}$ and $\vec{x}$ in the EF, while Latin
indices label the components of $\vec{a}$ (even if $\vec{\zeta}$
and $\vec{a}$ belong to the same physical space, as shown in (\ref{eq:a.initial.zeta}),
we prefer to use different indices for components of dynamical variables
and of the label $\vec{a}$).

The Hamiltonian (\ref{eq:H.EV}) is easily re-written in the LF as
\begin{equation}
H\left[\vec{\zeta},\vec{\pi},s\right]={\displaystyle \int_{\mathbb{D}_{0}}}d^{3}a\left[\frac{\pi^{2}}{2\rho_{0}}+\rho_{0}U\left(\frac{\rho_{0}}{\mathcal{J}},s\right)+\rho_{0}\varphi\left(\vec{\zeta}\right)\right].\label{eq:H.LV}
\end{equation}
$\rho_{0}\left(\vec{a}\right)$ is the initial mass density of the
$\vec{a}$-th parcel. The mass-specific internal energy density $U$
depends on the density of the parcel, that reads $\rho=\rho_{0}\mathcal{J}^{-1}$
because of mass conservation \cite{Pdhye.Thesis}, and on its entropy.
The dynamics of the non-dissipative limit in LV is governed by an
\emph{apparently canonical} Poisson bracket, reading:
\begin{equation}
\left\{ \Phi,\Psi\right\} _{L}={\displaystyle \int_{\mathbb{D}_{0}}}d^{3}a\left[\frac{\delta\Phi}{\delta\zeta^{\alpha}\left(\vec{a}\right)}\frac{\delta\Psi}{\delta\pi_{\alpha}\left(\vec{a}\right)}-\frac{\delta\Phi}{\delta\zeta^{\alpha}\left(\vec{a}\right)}\frac{\delta\Psi}{\delta\pi_{\alpha}\left(\vec{a}\right)}\right]\label{eq:Poisson.bracket.in.LV}
\end{equation}
(the expression ``apparently canonical'' will be commented later
on). For any physical observable $\Phi$ one has simply $\dot{\Phi}=\left\{ \Phi,H\right\} _{L}$,
giving rise to the equations of motion:
\begin{equation}
\begin{cases}
\dot{\zeta}^{\alpha}=\pi^{\alpha},\\
\dot{\pi}_{\alpha}=-\rho_{0}{\displaystyle \frac{\partial\varphi}{\partial\zeta^{\alpha}}}+A_{\alpha}\thinspace^{i}{\displaystyle \frac{\partial}{\partial a^{i}}}\left(\rho_{0}{\displaystyle \frac{\partial U}{\partial\mathcal{J}}}\right), & A_{\alpha}\thinspace^{i}={\displaystyle \frac{1}{2}}\epsilon_{\alpha\kappa\lambda}\epsilon^{imn}{\displaystyle \frac{\partial\zeta^{\kappa}}{\partial a^{m}}\frac{\partial\zeta^{\lambda}}{\partial a^{n}}},\\
\dot{s}=0
\end{cases}\label{eq:ideal.NS.LV}
\end{equation}
((\ref{eq:ideal.NS.LV}) the ``dot'' means ``time derivative along
the motion of the parcel'', also called \emph{Lagrangian}, or \emph{material},
\emph{derivative}).

In order to complete the dynamics of the non-ideal fluid in LF, the
metric part must be produced. The first step is to consider that the
equations of motion to be reproduced are the translation of the system
(\ref{eq:NS.viscosa.EV}) in parcel variables:
\begin{equation}
\begin{cases}
\dot{\zeta}^{\alpha}=\pi^{\alpha},\\
\\
\dot{\pi}_{\alpha}=-\rho_{0}{\displaystyle \frac{\partial\varphi}{\partial\zeta^{\alpha}}}+A_{\alpha}\thinspace^{i}{\displaystyle \frac{\partial}{\partial a^{i}}}\left(\rho_{0}{\displaystyle \frac{\partial U}{\partial\mathcal{J}}}\right)+\Lambda_{\alpha\eta\gamma\delta}\mathcal{J}\nabla^{\alpha}\nabla^{\gamma}\left({\displaystyle \frac{\pi^{\delta}}{\rho_{0}}}\right),\\
\\
\dot{s}=\dfrac{\mathcal{J}}{\rho_{0}T}\Lambda_{\alpha\beta\gamma\delta}\nabla^{\alpha}\left({\displaystyle \frac{\pi^{\beta}}{\rho_{0}}}\right)\nabla^{\gamma}\left({\displaystyle \frac{\pi^{\delta}}{\rho_{0}}}\right)+\dfrac{\kappa\mathcal{J}}{\rho_{0}T}\nabla^{\eta}\nabla_{\eta}T.
\end{cases}\label{eq:NS.viscosa.LV}
\end{equation}
The definition of $A_{\alpha}\thinspace^{i}$ was already given in
(\ref{eq:NS.viscosa.EV}). The operator $\nabla^{\mu}$ is the derivative
with respect to $\zeta_{\mu}$ intended as the differential operator
$\nabla^{\mu}=\frac{\partial a^{i}}{\partial\zeta_{\mu}}\frac{\partial}{\partial a^{i}}$,
and it acts on $\vec{a}$-dependent fields through the chain rule;
the operator $\nabla^{\mu}$ reads $\nabla^{\mu}=\left(J^{-1}\right)^{\mu i}\left(\partial\vec{\zeta}\right)\frac{\partial}{\partial a^{i}}$
in terms of the Jacobian $J\left(\partial\vec{\zeta}\right)$. In
(\ref{eq:NS.viscosa.LV}) $T$ represents the temperature of the $\vec{a}$-th
parcel

The metric bracket $\left(.,.\right)_{L}$ is obtained by requiring
that it reproduces the equations (\ref{eq:NS.viscosa.LV}) via the
prescription
\[
\dot{\Phi}=\left\{ \Phi,H\right\} _{L}+\lambda\left(\Phi,S\right)_{L}:
\]
in order to obtain it explicitly, one may consider $\left(\Phi,\Psi\right)_{E}$
in (\ref{eq:metric.bracket.EV}) and reason on the relationships between
the parcel variables and the Eulerian fields
\begin{equation}
\begin{cases}
\rho\left(\vec{x},t\right)={\displaystyle \int_{\mathbb{D}_{0}}}d^{3}a\rho_{0}\left(\vec{a}\right)\mathcal{J}\left(\partial\vec{\zeta}\left(\vec{a},t\right)\right)\delta^{3}\left(\vec{\zeta}\left(\vec{a},t\right)-\vec{x}\right),\\
\\
\vec{v}\left(\vec{x},t\right)={\displaystyle \int_{\mathbb{D}_{0}}}d^{3}a{\displaystyle \frac{\vec{\pi}\left(\vec{a},t\right)}{\rho_{0}\left(\vec{a}\right)}}\delta^{3}\left(\vec{\zeta}\left(\vec{a},t\right)-\vec{x}\right),\\
\\
\sigma\left(\vec{x},t\right)={\displaystyle \int_{\mathbb{D}_{0}}}d^{3}as\left(\vec{a},t\right)\delta^{3}\left(\vec{\zeta}\left(\vec{a},t\right)-\vec{x}\right).
\end{cases}\label{eq:L2E}
\end{equation}
The Eulerian field is the value taken by the corresponding Lagrangian
quantity attributed to the parcel that, \emph{at that given time},
transits \emph{at that given point}: hence, one should understand
$\rho_{0}\left(\vec{a}\right)\mathcal{J}\left(\partial\vec{\zeta}\left(\vec{a},t\right)\right)$
in (\ref{eq:metric.bracket.EV}) in the place of $\rho\left(\vec{x},t\right)$,
$\frac{\vec{\pi}\left(\vec{a},t\right)}{\rho_{0}\left(\vec{a}\right)}$
in the place of $\vec{v}\left(\vec{x},t\right)$ and $s\left(\vec{a},t\right)$
in the place of $\sigma\left(\vec{x},t\right)$, provided the label
$\vec{a}$ is chosen so that $\vec{\zeta}\left(\vec{a},t\right)=\vec{x}$.
The integral over $\mathbb{R}^{3}$ in $d^{3}x$ is replaced by an
integral over $\mathbb{D}_{0}$ in $d^{3}\zeta=\mathcal{J}d^{3}a$.

A special care must be used to treat the relationship between the
functional derivative with respect to any Eulerian field $\psi_{E}\left(\vec{x}\right)$
and that with respect to the corresponding Lagrangian variable $\psi_{L}\left(\vec{a}\right)$.
These operations are in fact defined via Frechet derivatives
\[
\begin{cases}
\dfrac{\delta\Phi}{\delta\psi_{E}\left(\vec{x}\right)}={\displaystyle \lim_{\epsilon\rightarrow0}}{\displaystyle \frac{d}{d\epsilon}}\Phi\left[\psi_{E}\left(\vec{x}'\right)+\epsilon\delta^{3}\left(\vec{x}'-\vec{x}\right)\right],\\
\\
\dfrac{\delta\Phi}{\delta\psi_{L}\left(\vec{a}\right)}={\displaystyle \lim_{\epsilon\rightarrow0}}{\displaystyle \frac{d}{d\epsilon}}\Phi\left[\psi_{L}\left(\vec{a}'\right)+\epsilon\delta^{3}\left(\vec{a}'-\vec{a}\right)\right],
\end{cases}
\]
so that, even if $\psi_{E}$ and $\psi_{L}$ may be identified with
each other, still the distributions $\delta^{3}\left(\vec{a}'-\vec{a}\right)$
and $\delta^{3}\left(\vec{x}'-\vec{x}\right)$, here to be understood
as $\delta^{3}\left(\vec{\zeta}\left(\vec{a}'\right)-\vec{\zeta}\left(\vec{a}\right)\right)$,
do not exactly coincide: $\delta^{3}\left(\vec{a}'-\vec{a}\right)=\mathcal{J}\delta^{3}\left(\vec{\zeta}\left(\vec{a}'\right)-\vec{\zeta}\left(\vec{a}\right)\right)$.
As a result, one may write:
\[
\dfrac{\delta\Phi}{\delta\psi_{E}\left(\vec{\zeta}\left(\vec{a}\right)\right)}=\frac{1}{\mathcal{J}\left(\partial\vec{\zeta}\left(\vec{a}\right)\right)}\dfrac{\delta\Phi}{\delta\psi_{L}\left(\vec{a}\right)}.
\]

All in all, \emph{the metric bracket for a viscous fluid in LF} reproducing
equations (\ref{eq:NS.viscosa.LV}) with $S\left[s\right]$ as a metric
generator reads:
\begin{equation}
\begin{array}{l}
\left(\Phi,\Psi\right)_{L}=\\
\\
=\dfrac{1}{\lambda}\int\limits _{\mathbb{D}_{0}}\mathcal{J}d^{3}a\left\{ T\Lambda_{\alpha\beta\gamma\delta}\left[\nabla^{\alpha}\left(\dfrac{\delta\Phi}{\delta\pi_{\beta}}\right)-\dfrac{1}{\rho_{0}T}\nabla^{\alpha}\left({\displaystyle \frac{\pi^{\beta}}{\rho_{0}}}\right)\dfrac{\delta\Phi}{\delta s}\right]\left[\nabla^{\gamma}\left(\dfrac{\delta\Psi}{\delta\pi_{\delta}}\right)-\dfrac{1}{\rho_{0}T}\nabla^{\gamma}\left({\displaystyle \frac{\pi^{\delta}}{\rho_{0}}}\right)\dfrac{\delta\Psi}{\delta s}\right]+\right.\\
\\
\left.+\kappa T^{2}\nabla^{\eta}\left(\dfrac{1}{\rho_{0}T}\dfrac{\delta\Phi}{\delta s}\right)\nabla_{\eta}\left(\dfrac{1}{\rho_{0}T}\dfrac{\delta\Psi}{\delta s}\right)\right\} .
\end{array}\label{eq:metric.bracket.LV}
\end{equation}
The bracket (\ref{eq:metric.bracket.LV}) is easily shown to exhibit
all the necessary properties for it to be a metric bracket: it's thoroughly
symmetric in the $\Phi\leftrightarrow\Psi$ exchange, while about
semidefiniteness one may note
\[
\left(\Phi,\Psi\right)_{E}=\left(\Phi,\Psi\right)_{L}
\]
provided the correct ``dictionary'' is used, so that one may count
of the fact that $\left(\Phi,\Psi\right)_{L}$ inherits all the good
properties from those demonstrated for $\left(\Phi,\Psi\right)_{E}$,
in \cite{Morrison.84.001,Mate.Tassi.metriplectic.MHD} and references
therein.

With the finding (\ref{eq:metric.bracket.LV}) we have the complete
metriplectic algebra of viscous fluid dynamics in the LF, that can
be reported as:
\[
\begin{cases}
\dot{\Phi}=\left\langle \left\langle \Phi,F\right\rangle \right\rangle _{L},\\
\\
\left\langle \left\langle \Phi,\Psi\right\rangle \right\rangle _{L}=\left\{ \Phi,\Psi\right\} _{L}+\left(\Phi,\Psi\right)_{L},\\
\\
F=H+\lambda S,\\
\\
H={\displaystyle \int_{\mathbb{D}_{0}}}d^{3}a\left(\frac{\pi^{2}}{2\rho_{0}}+\rho_{0}U+\rho_{0}\varphi\right), & S={\displaystyle \int_{\mathbb{D}_{0}}d^{3}a\rho_{0}s}.
\end{cases}
\]

As anticipated before, the advantage of looking at the metriplectic
fluid dynamics in the LF, instead of in the EF, is that a certain
subtlety about entropy is clarified, that has to do with the question
of it to be a CI of the theory.

\section{Entropy and the Casimir invariant condition\label{sec:Entropy-Casimir}}

Back to what described in Section \ref{eq:S.Casimir}, we speculate
here on the entropy of fluids \cite{Morrison.84.001,Mate.Tassi.metriplectic.MHD},
that appear as in-between the ``two ways of being a Casimir'' C1
and C2.

On the one hand, this $S$ clearly encodes information on the $\mu$DoF
of the continuum, while the fluid velocity describe a macroscopic
point of view of the system, as it happens in the C2 case. On the
other hand, Morrison and Padhye had algebraic reasons to show, in
\cite{PadMorr.01} and \cite{PadMorr.02}, that this $S$ belongs
to a family of quantities conserved, via a ``C1 mechanism'', out
of the reduction of the algebra (\ref{eq:Poisson.bracket.in.LV})
to the set $\mathcal{A}_{E}$ of quantities $\Theta\left[\vec{\zeta},\vec{\pi},s\right]$
so that $\left\{ \Theta,C_{1,2}\right\} _{L}=0$, that become the
physical quantities in the EF, and are invariant under \emph{parcel
relabeling transformations} (RT) \cite{Pdhye.Thesis}. Examining the
LF of the fluid, with the RT more clearly readable, the opinion of
the author here has become that the viscous fluid may be considered
on the same foot as those mentioned in \cite{Fish.th.PhD} and \cite{algebrizing.friction},
classifying its $S$ in the C2 case.

The RTs, on which the LF to EF reduction is based, are smooth invertible
maps $\vec{a}\mapsto\vec{a}'\left(\vec{a}\right)$ that leave the
Hamiltonian (\ref{eq:H.LV}) and the Eulerian fields (\ref{eq:L2E})
unchanged. The quantities acting as symplectic generators of such
RT via the bracket $\left\{ .,.\right\} _{L}$ must belong to one
of either the following families of functionals
\begin{equation}
\begin{cases}
C_{1}\left[\vec{\zeta},\vec{\pi},s\right]={\displaystyle \int_{\mathbb{D}_{0}}}d^{3}a\varepsilon\left(\vec{a}\right)Q_{s}\left(\vec{a}\right), & Q_{s}=\epsilon^{ijk}\frac{\partial\pi_{\alpha}}{\partial a^{i}}\frac{\partial\zeta^{\alpha}}{\partial a^{j}}\frac{\partial s}{\partial a^{k}},\\
\\
C_{2}\left[s\right]={\displaystyle \int_{\mathbb{D}_{0}}}d^{3}a\mathcal{W}\left(s\right),
\end{cases}\label{eq:RT.generators.L}
\end{equation}
where $\varepsilon$ and $\mathcal{W}$ are arbitrary functions. The
quantity $Q_{s}$ is referred to as \emph{potential vorticity}, while
the entropy of the fluid is an example of $C_{2}$, with $\mathcal{W}=\rho_{0}s$.
Clearly both $C_{1}$ and $C_{2}$ are in involution with any quantity
in $\mathcal{A}_{E}$, so that if the reduction with respect to the
symmetry they generate is performed, they do become CI. The point
is that, due to the fact that no derivative with respect to $s$ appears
in $\left\{ .,.\right\} _{L}$, the quantities $C_{2}$ were ``already
CI'' in the symplectic algebra of the Lagrangian Formulation. Instead,
a non-trivial set exists of LF functionals $\Xi\left[\vec{\zeta},\vec{\pi},s\right]$
so that $\left\{ \Xi,C_{1}\right\} _{L}\ne0$, with $C_{1}$ given
in (\ref{eq:RT.generators.L}): this is the set of the quantities
that can be constructed in the LF but \emph{that have not a corresponding
Eulerian quantity}, because \emph{they are not RT-invariant} \cite{PadMorr.01,PadMorr.02,Pdhye.Thesis}.
Moreover, the Poisson bracket$\left\{ .,.\right\} _{L}$ has been
indicated as ``apparently canonical'' because, even if the Frechet
derivatives $\frac{\delta}{\delta\zeta^{\alpha}\left(\vec{a}\right)}$
and $\frac{\delta}{\delta\pi_{\alpha}\left(\vec{a}\right)}$ in (\ref{eq:Poisson.bracket.in.LV})
appear just like they would be expected to in canonical brackets,
still $s$ has no involvement in it, but is part of $\mathbb{V}_{L}$.
This means that the symplectic operator giving rise to $\left\{ .,.\right\} _{L}$
is \emph{degenerate on $\mathbb{V}_{L}$}, and the bracket is not
``properly'' canonical. It admits \emph{a nontrivial null space},
the set of the quantities $C_{2}$ in (\ref{eq:RT.generators.L})
of anything depending on $s$ only: we could visualize this by expressing
the matrix related to $\left\{ .,.\right\} _{L}$ as
\[
\begin{array}{cc}
\left\{ \Phi,\Psi\right\} =\left(\partial_{\psi}\Phi\right)^{\mathrm{T}}\cdot Z\cdot\partial_{\psi}\Psi, & Z=\left(\begin{array}{ccc}
0 & \boldsymbol{1}_{3} & 0\\
-\boldsymbol{1}_{3} & 0 & 0\\
0 & 0 & 0
\end{array}\right)\end{array}
\]
being $\psi=\left(\vec{\zeta},\vec{\pi},s\right)$, while one also
has $\mathbb{V}_{L}=\mathbb{R}^{6}\oplus\mathbb{R}$, being $\mathbb{R}^{6}$
that of canonical variables $\left(\vec{\zeta},\vec{\pi}\right)$,
and $\mathbb{R}$ that of $s$.

The physical difference between $S$ and any $C_{1}$ is that $S$
includes only the $\mu$DoF responsible for dissipation, while the
$C_{1}$s mix them with the centre-of-mass variables $\left(\vec{\zeta},\vec{\pi}\right)$.
In few words, only $S$ is expected to play the driving role in dissipation
processes.

The physical difference of roles for $C_{1}$ and $C_{2}$ in (\ref{eq:RT.generators.L})
persists in the metriplectic algebra of the fluid in the EF. In terms
of Eulerian fields those quantities appear as follows
\begin{equation}
\begin{cases}
C_{1E}\left[\rho,\vec{v},\sigma\right]={\displaystyle \int_{\mathbb{D}}}d^{3}x\rho\mathcal{C}_{1}\left(Q_{\sigma E}\right), & Q_{\sigma E}={\displaystyle \frac{1}{\rho}}\vec{\partial}\sigma\cdot\left(\vec{\partial}\times\vec{v}\right),\\
\\
C_{2E}\left[\rho,\sigma\right]={\displaystyle \int_{\mathbb{D}}}d^{3}x\rho\mathcal{C}_{2}\left(\sigma\right)
\end{cases}\label{eq:Casimir.EV}
\end{equation}
(use has been made of the symbol $\vec{\partial}=\frac{\partial}{\partial\vec{x}}$):
clearly, all the quantities $C_{1E}$ or $C_{2E}$ in (\ref{eq:Casimir.EV})
satisfy the prescription (\ref{eq:S.Casimir}), so one could be tempted
to generalize the expression of the free energy as
\[
\mathcal{F}=H+\lambda_{1}C_{1E}+\lambda_{2}C_{2E};
\]
the point is whether this gives rise to any sensible dynamics through
the metriplectic algebra $\left\langle \left\langle .,.\right\rangle \right\rangle _{E}=\left\{ .,.\right\} _{E}+\left(.,.\right)_{E}$;
for sure, \emph{as long as the metric bracket (\ref{eq:metric.bracket.EV})
is used}, the equations of motion (\ref{eq:NS.viscosa.EV}) are produced
only choosing $\lambda_{1}=0$ and $C_{2E}=S$, so that entropy seems
to play a role that no other CI plays: assuming (\ref{eq:NS.viscosa.EV}),
the evolution of any $C_{kE}$ may be expressed as $\dot{C}_{kE}=\lambda\left(C_{kE},S\right)_{E}$,
for $k=1,2$. The Casimir $C_{1E}$ instead does not generate any
dynamics.

\section{Conclusions\label{sec:Conclusions}}

A viscous fluid with suitable border conditions relaxes to an asymptotic
equilibrium due to the presence of dissipation, while it can be written
in a Hamiltonian form in its frictionless limit. This is a perfect
system to be put in a metriplectic form according to the prescriptions
of \cite{Morrison.bracket.formulation.84}, \cite{algebrizing.friction}
and references therein.

Fluids may be represented in EF or in LF, and the metriplectic framework
for the EF was already known \cite{Morrison.84.001}. Here, the metriplectic
algebra in the LF is obtained, adopting the parcel variables as in
\cite{Bennett.book} to describe the fluid in a metriplectic form:
the resulting picture is rather clearer than the one in EF.

The position of the center-of-mass of the $\vec{a}$-th parcel $\vec{\zeta}\left(\vec{a}\right)$
and its momentum $\vec{\pi}\left(\vec{a}\right)$ undergo the dissipative
interaction with the $\mu$DoF of the nearby parcels, encoded in the
entropy of nearby parcels (of course, $\vec{\zeta}\left(\vec{a}\right)$
and $\vec{\pi}\left(\vec{a}\right)$ cannot interact directly with
the $\mu$DoF \emph{of their own parcel}, since no internal force
can alter the motion of the centre-of-mass \cite{FeynmanLectures}).
The novel result is the expression (\ref{eq:metric.bracket.LV}) of
\emph{the metric bracket in parcel variables}, through which the metric
generator of dissipation, namely the fluid entropy $S$, makes viscosity
act.

The pure Hamiltonian limit of the metriplectic system would actively
involve only the variables $\vec{\zeta}\left(\vec{a}\right)$ and
$\vec{\pi}\left(\vec{a}\right)$, as demonstrated by the expression
(\ref{eq:Poisson.bracket.in.LV}), in which no derivative appears
with respect to fields encoding the $\mu$DoF. This renders the fluid
entropy $S$ a Casimir invariant ``of C2 type'': the degrees of
freedom encoded in $S$ act as ``external variables'' with respect
to the field configuration which would be sufficient to describe the
ideal fluid in LF, i.e. the Poisson algebra based on $\vec{\zeta}\left(\vec{a}\right)$
and $\vec{\pi}\left(\vec{a}\right)$. Hence, the metric generation
of dissipation in this case shows the same mechanism as presented
in \cite{algebrizing.friction} and in Chapter 8 of \cite{Fish.th.PhD}.

Despite the Lagrangian Formulation leads to equations of motion that
are more complicated than the ones in EF, stating the dynamics of
a viscous fluid in parcel variables appears crucial in order to describe
more transparently\emph{ coherent structures of matter} \cite{Chang.coherent.structures}.

In their EF, fluids (and plasmas) appear to be often characterised
by modes representing local subsets of the continuum in which the
parcels move with macroscopic scale correlations (e.g., in vortices
or current structures); collective variables describing such field
configurations will probably be better described by adopting parcel
variables $\left(\vec{\zeta},\vec{\pi},s\right)$, because long range
correlation are likely to form well defined patterns ``in the $\vec{a}$-space''
rather than ``in the $\vec{x}$-space'', since the ``$\vec{a}$-space''
is the set $\mathbb{D}_{0}$ of parcels' identities, where it is possible
to keep track of which parcel has interacted with which other one,
and hence developed correlation at mesoscopic scales.

Forthcoming studies will investigate the application of what obtained
here to the LF of vortices \cite{tetrault}, while a contact with
the tetrad formalism, describing parcels of various scales \cite{tetrad.vieillefosse,tetrad.cantwell,tetrad.chertkov},
will be made.

Last but not least, the LF of an MHD collisional plasma will be constructed,
as an extension of the present study to electromagnetic degrees of
freedom.\medskip{}
\medskip{}
\medskip{}

\begin{center}
\noun{Acknowledgements}
\par\end{center}

The author wants to acknowledge the contribution to his speculations
by P. J. Morrison, with whom he fruitfully discussed during his short
term missions at the ``Centre de Physique Theorique'' in Marseille,
France. There, he also had the opportunity of exchanging ideas with
E. Tassi, who is acknowledged as well.

\end{document}